\documentclass[toc]{PoS}
\usepackage{graphicx}
\usepackage{amssymb}
\usepackage{fontenc}
\usepackage{times}
\usepackage{mathptmx}

\newcommand{\be}{\begin{equation}}
\newcommand{\ee}{\end{equation}}
\newcommand{\bea}{\begin{eqnarray}}
\newcommand{\eea}{\end{eqnarray}} \newcommand{\nn}{\nonumber}

\def\slash#1{\setbox0=\hbox{$#1$}#1\hskip-\wd0\dimen0=5pt\advance
       \dimen0 by-\ht0\advance\dimen0 by\dp0\lower0.5\dimen0\hbox
         to\wd0{\hss\sl/\/\hss}}

\title{Majorana and the Infinite Component Wave Equations}

\ShortTitle{Majorana and the Infinite Component...}

\author{{Roberto Casalbuoni}\\
        Dipartimento di Fisica dell'Universita' di Firenze and Sezione INFN\\
        Via G. Sansone 1, 50019 Sesto Fiorentino (FI), Italy\\
        Galileo Galilei Institute for Theoretical Physics\\
        L.go E. Fermi 2, 50125 Firenze, Italy\\
        E-mail: \email{casalbuoni@fi.infn.it}}

\abstract{I review the paper of Majorana about relativistic
particles with arbitrary spin written in 1932. The main motivation
for this papers was the dissatisfaction about the negative energy
solutions of the Dirac equation. As such, the paper became
immediately obsolete due to the almost contemporaneous discovery of
the positron. However, for the first time, the unitary
representations of the Lorentz group were introduced. Majorana
considered two particular representations (named, after him,
Majorana representations) which enjoy many interesting properties. A
discussion about the reasons for its revival in the 60's is
presented.}
\FullConference{International Conference --- Ettore Majorana's legacy and the Physics of the XXI century ---\\
         October 5-6 2006\\
         University of Catania, Italy}

\begin{document}

\section{Introduction}

In 1932 Ettore Majorana published a paper, in italian,  by the title
"Relativistic theory of particles with arbitrary angular momentum"
\cite{majorana:32ab}. As it is well known the physical
interpretation of the Dirac equation \cite{dirac:28ab} was rather
problematic due the existence of negative energy solutions. In 1931
Dirac  proposed a solution in terms of the hole-theory
\cite{dirac:31ab} introducing a new kind of particles with the same
mass of the electrons and opposite charge, the positrons. The
positron was discovered by Anderson \cite{anderson:32ab} at the end
of 1931, and the paper with the first picture of a positron appeared
at the very beginning of 1932. It is not clear when Majorana wrote
this paper (probably during the summer, according to Amaldi) and in
which month of 1932 the paper appeared in Il Nuovo Cimento. However
it seems that the news of the discovery of the positron arrived in
Rome only around the end of 1932. So when Majorana conceived his
paper the problem of the negative energy states was still in his
mind. Therefore the aim of the paper was to construct a Dirac-like
equation with only positive energy solutions.  Majorana found  that
this is indeed possible, but that it is necessary that the wave
function transforms under unitary representations (UR) of the
homogeneous Lorentz group. These representations are infinite
dimensional, as  he discovered. The UR's were completely unknown at
that time, not only among physicists but also among mathematicians.
Majorana showed here his great mathematical ability and his
mastering of group theory finding out two simple unitary
representations for the wave function. I will discuss some of the
reason why his paper was  ignored at the time it was published,
although it would have been of great interest from the point of view
of group theory. Majorana's paper was recovered from the general
ignorance thanks to  Fradkin  in 1966 that, stimulated by Amaldi
whereas he was visiting  Rome, published a comment to the Majorana's
paper \cite{fradkin:66ab}. There were several reasons why this paper
was of interest to physicists in that period. I will mention the
incoming use of dynamical groups and the problem of saturating the
algebra of currents at $p=\infty$ with a set of single particle
states. I will discuss  why the equations of Majorana type,
involving unitary representations of the Lorentz group, were of
interest for the previous topics. Another interesting point of the
infinite component wave equations is that the CPT theorem does not
generally hold. In fact the proof of the theorem in relativistic
local field theories is valid only under the assumption that it is
possible to perform an analytical continuation in the parameters of
the Lorentz group. This condition is  satisfied in the case of
finite dimensional representations of the Lorentz group but it is
not for UR's. I will discuss briefly this point.

\section{The paper of Majorana about relativistic particles of arbitrary spin}

In the paper of Majorana   the following linear wave equation of the
Dirac type was introduced: \be (E+\vec\alpha\cdot\vec p-\beta
M)\psi=0\ee Since Majorana wanted to avoid negative energy
solutions, he required  $\beta$ to be a positive definite operator.
The other important point was that he did  not require the validity
of the Klein-Gordon equation, that is he did not ask for  a single
mass value associated to the wave function. As a consequence of the
positivity of the operator $\beta$ it follows that the wave function
must transform according to a UR of the Lorentz group. The argument
of Majorana is very simple and it is based on writing down an action
from which to derive the wave equation in a variational way. The
action is\be \int d^4x\,\psi^\dagger\left(E+\vec\alpha\cdot\vec
p-\beta M\right)\psi\ee Since $\beta$ is required to be positive
definite one can redefine the wave function according to \be
\tilde\psi=\beta^{1/2}\psi\ee The action becomes \be
 \int d^4x\,\tilde\psi^\dagger\left(\Gamma_0 E+
\vec\Gamma\cdot\vec p-M\right)\tilde\psi\ee where \be
\Gamma_0=\beta^{-1},
~~~\vec\Gamma=\beta^{-1/2}\vec\alpha\beta^{-1/2}\ee From this one
gets the wave equation \be \left(\Gamma_\mu
p^\mu-M\right)\tilde\psi=0,~~~\Gamma_\mu=(\Gamma_0,\vec\Gamma),~~~p^\mu=(E,\vec
p)\label{eq:2.6}\ee Since the action must be Lorentz invariant it
follow that the same must be true for
$\tilde\psi^\dagger\tilde\psi$. Therefore, under a Lorentz
transformation  $\tilde\psi$ must transform as a UR of the Lorentz
group\be \tilde\psi^\prime=S\tilde\psi, ~~~~S^\dagger S=1\ee

The next step made by Majorana was to evaluate the commutation
relations for the generators of the Lorentz group, which in modern
notations read \be [J_i,J_j]=i\epsilon_{ijk} J_k,~~~
[J_i,N_j]=i\epsilon_{ijk} N_k,~~~ [N_i,N_j]=-i\epsilon_{ijk} J_k\ee
where $\vec J$ are the generators of the  rotation group and $\vec
N$  the boost generators. The relation with the covariant generators
$J_{\mu\nu}$ is: \be J_k=\frac 1 2\,\epsilon_{kij}
J_{ij},~~~N_k=J_{k0}\ee Majorana was an expert in group theory. He
had in his bookshelf the Weyl's book on Quantum Mechanics
 and group theory \cite{weyl:28ab}, as well as other books more
mathematically oriented in the subject. In particular in Weyl's book
one can find the calculation of the matrix elements in the angular
momentum basis of the electric dipole operator. On the other hand,
according to Amaldi who knew the way of working of Majorana it is
very well possible that he did the calculation by himself.

Let me now pause for a while in the exam of the paper and let me
discuss a bit what we know today about the irreducible UR's of the
Lorentz group. First of all, in complete generality, since the
boosts are vector operators, their action on an angular momentum
basis, $|j,m\rangle$, can be written as\bea N_+ |j,m\rangle &=&
C_j[(j-m)(j-m-1)]^{1/2}|j-1,m+1\rangle-
A_j[(j-m)(j+m+1)]^{1/2}|j,m+1\rangle\cr&+&
C_{j+1}[(j+m+1)(j+m+2)]^{1/2}|j+1,m+1\rangle\cr N_- |j,m\rangle &=&
-C_j[(j+m)(j+m-1)]^{1/2}|j-1,m-1\rangle-
A_j[(j+m)(j-m+1)]^{1/2}|j,m-1\rangle\cr&-&
C_{j+1}[(j-m+1)(j-m+2)]^{1/2}|j+1,m-1\rangle\cr N_3 |j,m\rangle &=&
C_j[(j-m)(j+m)]^{1/2}|j-1,m\rangle- A_j m|j,m\rangle\cr&-&
C_{j+1}[(j+m+1)(j-m+1)]^{1/2}|j+1,m\rangle
 \label{eq:2.10}\eea
 where
 \be
 A_j=\frac{ij_0j_1}{j(j+1)},~~~C_j=\frac
 ij\left[\frac{(j^2-j_0^2)(j^2-j_1^2)}{4j^2-1}\right]^{1/2}\ee
These matrix elements depend on the pair $(j_0,j_1)$. These numbers
characterize the Casimir operators of the Lorentz group \bea
C_1&=&\frac 12 J_{\mu\nu}J^{\mu\nu}=\vec J^2-\vec
N^2=j_0^2+j_1^2-1\nn\cr C_2&=&\frac 1 4
\epsilon_{\mu\nu\rho\sigma}J_{\mu\nu}J_{\rho\sigma}= 2\vec
J\cdot\vec N=2i j_0 j_1\eea where we have used the matrix elements
of the Lorentz generators. We see that $(j_0,j_1)$ and $(-j_0,-j_1)$
are equivalent representations, therefore we may choose $j_0$ to be
positive. It turns out that $j_0$ is the minimum angular momentum in
the representation ($C_{j_0}=0$). For finite dimensional
representations $j_1=j_{\rm max}+1$ ($C_{j_{\rm max}+1}=0$). On the
other hand in the infinite dimensional case the spin content of the
representation is $j_0,j_0+1,\cdots$. Notice also that, since under
parity $C_2$ changes sign, we have \be P:~~(j_0,j_1)\Rightarrow
(j_0,-j_1)\ee So far about a generic irreducible representation. The
unitarity condition selects two series of representations:
\begin{itemize}
\item {\bf Principal series}: $j_0$ integer or half-integer,
$j_1$ pure imaginary.
\item {\bf Supplementary series}: $j_0=0$, $j_1$ real with
$|j_1|<1$.
\end{itemize}

Going back to the paper of Majorana, after the equations defining
the commutation relations among the Lorentz group generators he
wrote: "The simplest solutions [of the commutation relations] by
means of hermitian operators is given by the following infinite
matrices" and he writes down the matrix elements of the Lorentz
generators that can be obtained from the equations (\ref{eq:2.10})
by the choice $(j_0=0,j_1=1/2)$ or $(j_0=1/2,j_1=0)$. These two
representations, later named after him, correspond to $C_1=-3/4$ and
$C_2=0$ and are parity invariant. Majorana did not spend any other
word about this simplicity, so it is not clear what he meant by it.
One possibility is that he refers to the fact that with this choice
the matrix elements of the boosts are particularly simple, since the
coupling $j\to j$ vanishes and the other two couplings $j\to j\pm 1$
are given by $i/2$. Since in his paper Majorana mentioned the fact
that for his two representations $C_2=0$, it is also possible that
he realized that with his choice the theory is parity invariant, but
he did not state  it explicitly. Also, quite strangely, he did not
mention the other invariant operator $C_1$.

At this point Majorana went on to the evaluation of the matrix
elements of the four-vector operator $\Gamma_\mu$. He wrote  the
commutation relations with the Lorentz generators that, using again
modern notations, read \be
[J_{\mu\nu},\Gamma_\rho]=i\left(\Gamma_\mu g_{\nu\rho}- \Gamma_\nu
g_{\mu\rho}\right)\ee Then, Majorana wrote directly the matrix
elements of $\Gamma_\mu$. These can be obtained by the observation
that once $\Gamma_0$ is known, the $\Gamma_i's$ can be evaluated
from\be \Gamma_i=-i[N_i,\Gamma_0]\ee Noticing that $\Gamma_0$ is a
scalar under rotations \be \langle j',m'|\Gamma_0|j,m\rangle=\langle
j,m|\Gamma_0|j,m\rangle\delta_{jj'}\delta{mm'}\ee and using \be
\Gamma_0=[[N_i,\Gamma_0],N_i]\ee he got $\Gamma_0$ up to a constant.
Choosing the constant to be one: \be \langle
j,m|\Gamma_0|j,m\rangle=j+1/2\ee Going to the rest frame (for the
moment being we are considering time-like solutions) Majorana found
the mass spectrum \be M_j=\frac M{j+1/2},~~j=j_0, j_0+1,\cdots,~~j_0
=0~{\rm or}~1/2\ee By reading the Majorana paper there is no trace
about the fact that in order to write down a relativistic equation
involving a four-vector operator one has to require very stringent
constraints on the representation chosen . As we shall see, the two
Majorana representations are the only irreducible representations
for which these constraints are satisfied.

\subsection{Four-vector operators}

We take another pause in the analysis of  Majorana's paper to study
the problem of a relativistic wave equation of the Dirac (or
Majorana) type involving a four-vector operator. The consistency of
the wave equation \be p\cdot\Gamma\psi=M\psi\ee requires that the
representation to which $\psi$ belongs  is contained in the tensor
product of the four-vector representation with the representation
itself. Symbolically \be
\psi\subset\Gamma_\mu\otimes\psi\label{eq:2.19}\ee In order to
evaluate this tensor product, let me first consider the case of
finite dimensional representations. Remember that the Lorentz group
is isomorphic to $SU(2)\otimes SU(2)$ and correspondingly a
finite-dimensional representation can be denoted by $(s_1,s_2)$
where $s_1$ and $s_2$ are the spin content of the two $SU(2)$
representations. The physical spin operator is the sum of the two
spin of the two commuting groups $SU(2)\otimes SU(2)$. The relation
with the notation $(j_0,j_1)$ is \be
j_0=|s_1-s_2|,~~j_1=(s_1+s_2+1)sign(s_1-s_2)\ee The spin content of
a  four-vector is 0 and 1. Therefore we have \be \Gamma_\mu\in
(j_0,j_1)=(0,2) \Rightarrow (s_1,s_2)=(1/2,1/2)\ee To evaluate the
direct product of two finite dimensional representations of the
Lorentz group is trivial in the basis $(s_1,s_2)$, since we have
only to combine separately the spin of the two representations of
$SU(2)$. We get \bea \left(\frac 12,\frac
12\right)\otimes(s_1,s_2)&=&\left(s_1+ \frac 12,s_2+\frac 12\right)
\oplus\left(s_1+\frac 1 2,s_2-\frac 12\right)\oplus\nn\cr
&\oplus&\left(s_1-\frac 12,s_2+\frac 12\right)\oplus\left(s_1- \frac
12,s_2-\frac 12\right)\eea Going back to the $(j_0,j_1)$ notations
we obtain\be (0,2)\otimes(j_0,j_1)=(j_0,j_1+1)\oplus(j_0+1,j_1)
\oplus(j_0-1,j_1)\oplus(j_0,j_1-1)\ee One can show that this
relation holds for any irreducible representation $(j_0,j_1)$. Since
the condition in eq. (\ref{eq:2.19}) reads \be \pm(j_0,j_1)\subset
(0,2)\otimes(j_0,j_1)\ee we find easily that the only solutions to
this condition are indeed the two Majorana representations \be
(j_0,j_1)=(0,1/2)~~{\rm or}~~(1/2,0)\ee

More solutions can be found by relaxing the condition that $\psi$
belongs to an irreducible representation. Consider the case of
"coupled representations". This means that \be \psi\in
(j_0,j_1)\oplus(j_0^{\,\prime},j_1^{\,\prime})\ee such that
\be(j_0,j_1)\subset(0,2)\otimes(j_0^{\,\prime},j_1^{\,\prime}),~~~
(j_0^{\,\prime},j_1^{\,\prime})\subset(0,2)\otimes(j_0,j_1)\ee In
this way we satisfy the constraint of eq. (\ref{eq:2.19}). For
instance, the Dirac representation \be(j_0,j_1)=\left(\frac 12,\frac
32\right)\oplus\left(\frac 12,-\frac 32\right)\approx\left(\frac 12,
0\right)\oplus\left(0,\frac 12\right)=(s_1,s_2)\ee satisfies this
condition. If we are in the case $M=0$ there is no such constraint
on the wave equation \be p^\mu\Gamma_\mu\psi=0\ee and we get, for
example, the Weyl's equations for massless spinors $(0,1/2)$ and
$(1/2,0)$. For coupled representation one gets the Majorana equation
for massive neutral particles \cite{Majorana:ab1937}, by observing
that under hermitian conjugation, in a finite dimensional
representation $C_2\to -C_2$ and therefore $j_1\to -j_1$ or
equivalently $(s_1,s_2)\to (s_2,s_1)$. In particular $(1/2,0)$ is
equivalent to $(1/2,0)^*$. More precisely \be i\sigma_2 \chi^*\in
(0,1/2),~~~{\rm if}~ \chi\in (1/2,0)\ee As a consequence, it is
possible to define a 4-component Majorana spinor by putting together
$\chi$ and $\chi^*$ \be
 \psi_M=\left(\matrix{\chi\cr i\sigma_2\chi*}\right)\ee
This spinor satisfies the condition of (pseudo-) reality \be
C\psi_M= \left(\matrix{ 0 & -i\sigma_2\cr -i\sigma_2 & 0}\right)
\left(\matrix{\chi\cr i\sigma_2\chi*}\right)= \left(\matrix{\chi*\cr
-i\sigma_2\chi}\right) =\psi_M^*\ee

An interesting question is if Majorana was aware of the constraints
on the representation in order to write a linear wave equation and
in the affirmative case if this helped him in finding out  the
equation for neutral massive particles  beyond the infinite
component wave equation.

\subsection{Other interesting points discussed by Majorana}

We have seen that the Majorana equation gives rise to a mass
spectrum for particles with different mass and spin. At that time
this was not very interesting, since the spectrum of known particles
was very poor (essentially, $p$, $e$, $\gamma$ and perhaps $n$).
Therefore the idea of Majorana was rather to get the formalism to
find the wave function for a particle with  given spin and mass.
This is something that it is possible to obtain from the Majorana
wave equation in the non relativistic limit. In fact he shows that
if one takes a solution of  the wave equation, $\psi_{s,m}$, with
fixed spin $s$, and mass $M/(s+1/2)$,  then the wave functions for
particles with different spin are suppressed by orders of $v/c$. For
instance \be \psi_{s-1,m}\approx {\cal O} \left(\frac v c\right),~~~
\psi_{s-2,m}\approx {\cal O} \left(\frac {v^2}{
c^2}\right),~~~\cdots\ee The proof goes like the decoupling of the
negative energy solutions of the Dirac equation in the limit $v/c\to
0$.

Another point analyzed by Majorana was the existence of space-like
solutions. For space-like momenta one goes to the special frame \be
p^\mu\to (0,0,0,\tilde p^3), ~~~p^2=-\left(\tilde{p}^{3}\right)^2\ee
The equation becomes \be \tilde p^3\Gamma_3\psi=M\psi\ee It is
possible to diagonalize simultaneously $\Gamma_3$ and $J_3$ \be
\Gamma_3|\sigma, m\rangle=\sigma|\sigma,
 m\rangle,~~~J_3|\sigma,m\rangle=m|\sigma, m\rangle\ee
with $\sigma\ge 0$. The eigenvalue of $\Gamma_3$ is connected with
the Casimir of the little group $SU(1,1)$ of the space-like
momentum. Therefore one gets a continuous of space-like solutions:
\be p_0^2-\vec p^{\,2}=-\frac{M^2}{\sigma^2}< 0\ee The Majorana
equation has also a continuous of light-like solutions, but we will
skip their discussion since they do not appear in the paper by
Majorana.

 Majorana   wave equation was rediscovered and generalized
(often without reference nor to the original paper
 neither to the existence of space-like solutions), as for instance in
 refs. \cite{gelfand:48ab, ginzburg:56ab} or in the books by Gel'fand, Minlos
 and Shapiro \cite{gelfand:63bc} and Naimark \cite{naimark:64ab}.

\section{Why the paper by Majorana was ignored?}

I will list here some of the topics that most probably contributed
to take this paper out of the mainstream of research at that time.

We have already discussed the fact that the positron was discovered
the same year of the Majorana equation and, presumably, the paper
lost soon its physical interest also to the  Majorana's eyes.

The Majorana wave equation gives rise to a mass spectrum, whereas at
his time only a very restricted number of particles were known.
Therefore this point did not make the theory very attractive, and
Majorana himself did not pay too much attention to it.

Group theory was not yet very popular among physicists. On the
contrary Majorana was very interested in this subject since his
graduation time. In fact in his notes a lot of space is dedicated to
group theoretical  calculations.

The paper was written in italian and on Il Nuovo Cimento. This
journal was not widely read at that time. Furthermore the Science
Abstract did not abstract from Nuovo Cimento until 1946. On the
other hand it was abstracted from Physikalische Berichte
 and the abstract of the Majorana's paper
was reviewed there \cite{physikalische:33ab}. Unfortunately the
abstractor was not an expert on the subject and the novelties of the
paper, as the first treatment of the unitary representations of the
Lorentz group, were not underlined.

In 1939 Wigner \cite{wigner:39ab} solved completely the problem of
the unitary representations of the Poincar\'e group, i.e. the
inhomogeneous Lorentz group. As it is well known physics requires
that the wave function transforms under a UR of the latter group.
These are obtained by fixing a representation of the translation
group, that is to say the four-momentum, $p^\mu$. Given $p^\mu$ one
has to choose a UR of the corresponding little group ($SU(2)$,
$SU(1,1)$ or $E2$ according to a time-like, space-like or light-like
momentum). This means that it is not mandatory to make use of UR's
of the Lorentz group. Indeed in the Dirac representation only the
angular momentum part is unitary, whereas the boost transformations
are not. In the 30's (and even later), before Wigner's contribution,
the situation was rather confused. This has lead Streater
\cite{streater:67ab}, commenting his own article
\cite{streater:67cd} about the problems of the infinite component
wave equations  to the following considerations about the Dirac
equation:

{\it I conjecture that Dirac had the mistaken belief that his
(Dirac) equation did not give rise to a unitary representation of
the inhomogeneous Lorentz group (the Poincaré group), because the
$4\times 4$ matrices appearing in it were not unitary. He may have
realised the great importance of unitary representations after
Wigner's book, Group Theory with applications to atomic
spectroscopy. It might be that the fear that his equation were badly
wrong urged Dirac to invent, in about 1945, single-handedly, some
irreducible unitary representations of the Lorentz group, a task
thought to be too hard for mathematicians at the time. If so, it was
all to no avail, as the unitarity of the representation (of the
Poincaré group, as opposed to the Lorentz group acting on the
spinors) given by the original Dirac equation was shown by Wigner
(1939) and by Bargmann and Wigner (1947).}

We have already noticed that the Majorana wave equation was later
rediscovered, often without any reference to the original paper.
However in 1966, Fradkin, after a suggestion by Amaldi published a
paper in english \cite{fradkin:66ab} commenting the Majorana's
paper. In fact at that time two lines of research pointed toward
infinite component wave equations. These two topics were:
\begin{itemize}
\item Dynamical groups.
\item The Gell-Mann's program of saturating the algebra of currents
at $p=\infty$ in terms of single particle states.
\end{itemize}
I will discuss the reasons for this renewal of interest in the next
Section. I should also mention that the possibility of getting a
mass spectrum was rather interesting from the point of view of the
Regge theory.

\section{The new interest in the 60's for the infinite component
wave equations}

\subsection{Dynamical groups}

In the 60's the main roads to strong interactions were the
analytical $S$-matrix and  group theory. A very important result
obtained by group theory was the discovery of the symmetry $SU(3)$
as a classification group for hadrons and its unification with the
rotation group leading to the $SU(6)$ symmetry
\cite{Gursey:1992dc,Sakita:1964qq,Sakita:1964qr}. A justification of
the success of this group was  the idea that hadrons were composite
objects. In this sense it was natural to try to learn something from
the simplest known system, the hydrogen atom. The similarity of the
problems leading to $SU(6)$ for hadrons and to $O(4)$ for the
hydrogen atom, unifying the rotation group with an internal symmetry
($SU(3)$ for hadrons and the transformations generated by the
Runge-Lenz vector for the H-atom), was discussed in ref.
\cite{dothan:65ab}. In 1967, Barut and Kleinert \cite{barut:67ab}
and Fronsdal \cite{fronsdal:67ab} found that it was possible to
enlarge the symmetry group $O(4)$ of the H-atom to a dynamical group
$O(4,2)$ which included, among its generators, the electric dipole
operator. The name of this approach, "dynamical groups" was due to
the hope to be able to describe the interactions as generators of a
group, as it happens for the electromagnetic interaction in the
H-atom. It is interesting that the previous authors were able to
prove that the Schr\"odinger equation for the hydrogen, taking also
into account the electromagnetic interaction, can be rewritten in
the form of a non-relativistic infinite component wave equation of
the Majorana type. The connection between composite systems and UR's
of the Lorentz group was already known by Dirac \cite{dirac:45ab}
(see also the Streater's comment \cite{streater:67ab}) who discussed
the UR of this group in terms of continuous variables (instead of
using a discrete basis as in the Majorana's approach). Eventually
this idea of Dirac was the basis of the bilocal field theory
introduced by Yukawa \cite{yukawa:53ab,Yukawa:1953es}. On the basis
of these considerations several authors discussed a series of
infinite component relativistic wave equations
\cite{fronsdal:67cd,nambu:66ab,nambu:67ab}. However all these
equations showed a number of diseaeses as:
\begin{itemize}
\item Presence of redundant solutions, as the space-like ones.
\item Typically these equations  violate the CPT theorem (as
for the Majorana case where no negative energy solutions are
present).
\item The spin-statistics theorem does not generally hold
\cite{feldman:67ab,fronsdal:67ef,streater:67cd}.\end{itemize} As a
consequence of these problems the program of dynamical groups died
very rapidly.

\subsection{Current algebra}

Let me consider the matrix elements of a vector current (to fix the
ideas I will take currents of $SU(3)\otimes SU(3)$) in the limit of
external momenta going to infinity \cite{dashen:ab1966}. To this end
let me define \be F_i(\vec q)=\int d^3\vec x\, e^{i\vec q\cdot\vec
x} j_i^0(\vec x,0),~~~i\in SU(3)\otimes SU(3)\ee The matrix elements
of this operator at $p=\infty$ can be written as \be
\lim_{p_3,p_3'\to\infty}\langle\vec p',N|F_i(\vec q)|\vec p,
N\rangle=\delta^3(\vec {p'}+\vec q-\vec p) \left(N'|J_i(\underline
p'-\underline p)|N\right) \ee where ${\underline p}=(p_1,p_2)$. In
the previous equation the states $|N)$ are fictitious states that,
in the case of single particles, depend only on the quantum numbers
$N$. The algebra of the operators $J_i(\underline p)$ is
\be[J_i(\underline p),J_j(\underline p')]= if_{ijk}J_k(\underline
p+\underline p'),~~~ J_i^\dagger(\underline p)=J_i(-\underline p)\ee
The idea was to look for representations of this algebra in the
space of states of single particle \cite{gellmann:ab1966}.  As shown
by Coester and Roepstoff \cite{coester:67ab}, in the space of single
particle states, this algebra has only infinite dimensional
representations.

An obvious representation of this algebra is: \bea &J_i(\underline
p)=\sum_n\frac 12\lambda_i^{(n)} e^{i\underline p\cdot\underline
x^{(n)}},~~~[x_1^{(n)},x_2^{(m)}]=0&\cr\cr &n=1,2~ {\rm
for~mesons},~~~n=1,2,3~ {\rm for~baryons}&\label{eq:4.4}\eea
Unfortunately $J_i$ arises as a limit of the fourth component of a
four-vector operator and therefore there is a set of complicated
conditions that it must satisfy, the so called angular conditions
\cite{dashen:66ab}. This can be easily understood considering the
matrix element \be \langle \vec{p'},N'|j^\mu|\vec p,
N\rangle,~~~{\rm with}~j'=0,~j\ge 1\ee It depends on 4 invariant
form factors whereas, without requiring any other condition,
$\left(N'|J(\underline p)|N\right)$ depends on $2j+1$ form factors.
As I said the angular conditions are very complicated and just to
convince you  I will write their infinitesimal form \be
{\left\{I,\left\{ I,\left\{I,J_i(\underline
p)\right\}\right\}\right\}= \frac 1
4\left[M^2,\left[M^2,\left\{I,J_i(\underline
p)\right\}\right]\right]+} {+\frac 1 2 |\underline
p|^2\left[M^2,\left\{I,J_i(\underline p\right\}\right]+ \frac 1 4
|\underline p|^4\left\{I,J_i(\underline p)\right\}}\ee where
\be{\left\{I,J_i(\underline p)\right\}=\frac 12
\left[M^2,\left[L_3,J_i(\underline p)\right]\right]- \frac 12
|\underline p|^2\left[L_3,J_i(\underline p)\right]_+
-\left[\underline p\cdot M\underline L,J_i(\underline
p)\right]},~~~{\underline L=(L_1,L_2)}\ee Furthermore
\be{\left[L_3,J_i(\underline p)\right]=i\underline
p\wedge\underline\nabla_p J_i(\underline
p)},~~~{\underline\nabla_p=\left(\frac{\partial}{\partial
p_1},\frac{\partial} {\partial p_2}\right)}\ee When all the
particles have the same mass, $M$, the  exponential solution (see
eq. (\ref{eq:4.4})) satisfies all these conditions with a position
operator given by \be (x_1,x_2)=\frac 1 M(F_1,F_2)=\frac 1
M(N_1+J_2,N_2-J_1)\ee where $(F_1,F_2)$ are the $E2$ generators (the
little group of light-like momenta). Starting from the degenerate
case it is possible to find an approximate solution through a
perturbative expansion in the splitting mass term
\cite{gellmann:67az,gellmann:67ay}.

It is easy to see that one can avoid the problem of solving the
angular conditions by using an appropriate infinite component wave
equation. This is done by the following steps:
\begin{itemize}
\item Introduce a wave function, $\psi(x)$ transforming according to some UR of
the Lorentz group with an appropriate spin content.\item Write an
invariant wave equation with the desired mass spectrum: \be
D(x)\psi(x)=0\ee However notice that the mass squared should not
increase more than the angular momentum as shown in ref.
\cite{grodski:67yz} (an equation with a linear mass spectrum was
proposed in \cite{casalbuoni:ac1971}).
\item Require that the wave equation is invariant under a specified
internal symmetry group.\end{itemize} Using the last requirement one
can construct an  algebra of conserved currents satisfying
automatically current algebra as a consequence of the canonical
commutation relations among the fields.

Of course there are problems very similar to the ones listed for the
case of dynamical groups, but the real problem here (and also in the
other case) is the presence of the space-like solutions. Before
discussing further this point I want to stress that, immediately
after the formulation of the program of saturating the current
algebra, a No-Go theorem was formulated by Grodsky and Streater
\cite{grodsky:67ab}. These authors made the following assumptions:
\begin{enumerate}
\item existence of covariant wave functions,
\item reasonable mass spectrum: only time-like solutions with finite
degeneracy on each mass shell,
\item the representation of the wave function should contain at
least one finite-dimensional representation of the Lorentz group and
a four-vector operator should exist,
\item the solutions of the wave equation form a complete set,
\end{enumerate}
and proved that each mass-shell must be infinitely degenerate.
Therefore the set of solutions satisfying the previous assumptions
is void. On the other hand the theorem does not hold if the wave
equation has space-like solutions. There could be a way out if the
space-like solutions would decouple from the time-like ones, or said
in other words, if the space-like solutions would not contribute to
the completeness relation. It is possible to show that in these
theories the decoupling of the space-like solutions is related to
the possibility of proving  the CPT theorem (see the discussion in
ref. \cite{casalbuoni:73ab}). We shall see, in the next Section,
that it is not possible to prove the CPT theorem for UR's and
therefore the decoupling of space-like solutions cannot occur.

\section{The CPT theorem}

Consider a wave equation of the Dirac or Majorana type. The wave
operator will be CPT invariant if it possible to define the
following operation \be p^\mu\to -p^\mu,~~~\Gamma_\mu \to
-\Gamma_\mu\ee The transformation of $\Gamma_\mu$ can be obtained
through a rotation of $\pi$ along the third axis followed by a boost
along the same direction by an imaginary boost parameter $\xi=i\pi$:
\be R_3(\pi)\Gamma_{1,2}R_3(\pi)^{-1}=-\Gamma_{1,2},~~~
B_3(i\pi)\Gamma_{0,3}B_3(i\pi)^{-1}=-\Gamma_{0,3}\ee In the case of
the Dirac equation: \be
B_3(i\pi)R_3(\pi)=(\gamma_0\gamma_3)(\gamma_1\gamma_2)=
\gamma_0\gamma_1\gamma_2\gamma_3\approx\gamma_5\ee It should be
clear that in the case of Majorana such an operation cannot exist
since $\Gamma_0$ is positive definite and the equation has not
negative energy solutions. The reason why the operation does not
exist is that for all the irreducible UR's of the Lorentz group the
operator $B_3(\xi)$ has  a pole at $\xi= i\pi$. The same happens for
all the irreducible infinite dimensional represntations where the
infinitesimal boost generator, $N_3$, is  a normal operator. As a
consequence of this pole, the standard derivation of the CPT
theorem, as for instance in the Streater and Wightman book
\cite{streater:64ab} based on the analytical continuation of the
Lorentz group, does not hold. However, for all the finite
dimensional representations of the Lorentz group, the operator
$B_3(i\pi)$ exists and the CPT theorem is valid. It turns out that
also the spin-statistics theorem is based on the existence of such
an operator ($B_3(i\pi)$) \cite{streater:64ab}. Therefore also this
theorem cannot be proved in these theories. Of course  the
possibility of arranging the theory in such a way that CPT and
spin-statistics are satisfied is not excluded.

To stress the previous point let me notice that, also if the wave
equation has negative energy solutions, the CPT theorem can be
violated. Consider, for instance, the following wave equation
\cite{abers:67ab}: \be\left(p^\mu\gamma_\mu-M-\frac 12
\sigma^{\mu\nu}[\Gamma_\mu,\Gamma_\nu]\right)\psi=0\ee with $\psi$
transforming as the direct product of the representations
Dirac$\otimes$Majorana. The existence of negative energy solutions
is guaranteed by the CPT transformation \be p^\mu\to -p^\mu,
~~~\gamma_\mu\to -\gamma_\mu\ee Notice that under this
transformation $\Gamma_\mu\to\Gamma_\mu$. Consider now a vector
field $\phi^\mu$ with definite properties of transformations under
the the previous CPT operation. We may construct the two local
couplings:
\be\bar\psi\gamma_\mu\psi\phi^\mu,~~~~\bar\psi\Gamma_\mu\psi\phi^\mu\ee
Clearly one of these two couplings is not CPT invariant.

\section{Conclusions}

Nowdays the UR's of the Lorentz group do not seem to have
interesting physical applications. However, in the paper that we
have reviewed here, Majorana shows all his mathematical strength and
ingenuity. There are several interesting points raised up by this
paper. One is the question of the choice of representations made by
Majorana. They enjoy many properties, they are the only irreducible
UR's for which it is possible to define a four-vector operator in
the sense specified in Section 2. Furthermore they are parity
invariant. These properties can be easily seen by means of the group
theoretical analysis developed here, so the question is if Majorana
had these notions, or he arrived to these representations by simple
chance. Considering that applying the same considerations it is easy
to understand also the Majorana formulation of the massive neutral
particle it seems more incredible that all this derived by casual
circumstances.

 Also, quite interestingly,  the work of
Majorana shows clearly that the CPT theorem can be violated in a
local relativistic theory. In fact, relativity and locality are not
enough to ensure the validity of this theorem. A further hypothesis
about the nature of the representations of the Lorentz group is
necessary. In particular the CPT theorem is valid for any finite
dimensional representation. Therefore the only way to get a
reasonable theory for a mass spectrum must involve necessarily
finite dimensional representations (unless to introduce space-like
solutions,  violation of CPT and/ or lack of the spin-statistics
connection). This happens in  string theory which involves an
infinite number of finite dimensional representations. It is curious
that in the last case there are, in principle, problems with the
positivity of the states. As well known this problem  can be avoided
by choosing a particular value for the space-time dimensions. On the
contrary, in the case of unitary representations this problem does
not arise, the metrics in the space of the states is positive
definite, but of course, as we have seen, the theory has many other
difficulties.


\newpage



\begin{thebibliography}{99}
\bibitem{majorana:32ab}
E. Majorana, Nuovo Cimento {\bf 9} (1932) 335.

\bibitem{dirac:28ab}
P.A.M. Dirac, Proceedings of the Royal Society {\bf A117} (1928)
610.

\bibitem{dirac:31ab}
P.A.M. Dirac, Proceedings of the Royal Society {\bf A133} (1931)
610.

\bibitem{anderson:32ab}
C.D. Anderson, Science {\bf 76} (1932) 238.

\bibitem{fradkin:66ab}
D.M. Fradkin, American Journal of Physics {\bf 34} (1966) 314.


\bibitem{weyl:28ab}
H. Weyl, {\it Gruppentherie und Quantunmechanik}, S. Hierzel Verlag,
Leipzig (1928).

\bibitem{Majorana:ab1937}
E. Majorana, Il Nuovo Cimento, {\bf 14} (1937) 171.

\bibitem{gelfand:48ab}
Gel'fand and Yaglom, JETP {\bf 18} (1948) 703, 1096, 1105.

\bibitem{ginzburg:56ab}
V.L. Ginzburg, Acta Physica Polonica {\bf 15} (1956) 163.

\bibitem{gelfand:63bc}
I.M. Gel'fand, R.A. Minlos
 and Z. Ya. Shapiro, {\it Representations of the Rotation and Lorentz Group and their
Applications},
 Pergamon Press (1963).
\bibitem{naimark:64ab}
 M. Naimark, {\it Linear Representations of the Lorentz Group},
 Pergamon Press (1964).

\bibitem{physikalische:33ab}
Physikalische Berchte (1933) - {\bf I} 548.

\bibitem{wigner:39ab}
E. Wigner, Annals of Mathematics {\bf 40} (1939) 149.

\bibitem{streater:67ab}
R. F. Streater, http://www.mth.kcl.ac.uk/~streater/rongspin.html

\bibitem{streater:67cd}

R. F. Streater, Commun. in Mathematical Phys. {\bf 5} (1967) 88.

\bibitem{Gursey:1992dc}
  F.~Gursey and L.~A.~Radicati,
  Phys.\ Rev.\ Lett.\  {\bf 13} (1964) 173.

\bibitem{Sakita:1964qq}
  B.~Sakita,
  Phys.\ Rev.\  {\bf 136} (1964) B1756.

\bibitem{Sakita:1964qr}
  B.~Sakita,
  Phys.\ Rev.\ Lett.\  {\bf 13} (1964) 643.


  \bibitem{dothan:65ab}

  Y. Dothan, M. Gell-Mann and Y. Ne'eman, Physics Letters {\bf 17}
  (1965) 148.


  \bibitem{barut:67ab}
A.O. Barut and H. Kleinert, Physical Review {\bf 156} (1967) 1541.

\bibitem{fronsdal:67ab}
C. Fronsdal, Physical Review {\bf 156} (1967) 1665.

\bibitem{dirac:45ab}
P.A.M. Dirac, Proceedings of the Royal Society {\bf A183} (1945)
284.

\bibitem{yukawa:53ab}
H. Yukawa, Physical Review {\bf 77} (1953) 219

\bibitem{Yukawa:1953es}
  H.~Yukawa,
  Phys.\ Rev.\  {\bf 91} (1953) 415.
\bibitem{fronsdal:67cd}
C. Fronsdal, Physical Review {\bf 156} (1967) 1665.

\bibitem{nambu:66ab}
Y. Nambu, Supplements of Progress in Theoretical Physics {\bf 37}
and {\bf 38} (1966) 368.

\bibitem{nambu:67ab}
Y. Nambu, Physical Review {\bf 160} (1967) 1171.

\bibitem{feldman:67ab}
G. Feldman and P.T. Matthews, Physical Review {\bf 154} (1967) 1241.

\bibitem{fronsdal:67ef}
C. Fronsdal, Physical Review {\bf 154} (1967) 1241.

\bibitem{dashen:ab1966}

R.F. Dashen and M. Gell-Mann, {\it Algebra of current componets at
infinite momentum}, Proceedings 3$^rd$ Coral Gables Conference on
"Symmetry Principles at High Energy", W.H. Freeman Co., San
Francisco (1966).

\bibitem{gellmann:ab1966}

M. Gell-Mann, {\it Current algebra. Strong and weak interactions},
International School of Physics E. Majorana (1966).


\bibitem{coester:67ab}

F. Coester and G. Roepstoff, Physical Review {\bf 155} (1967) 1583.

\bibitem{dashen:66ab}
R.F. Dashen and M. Gell-Mann, Physical Review Letters {\bf 17}
(1966) 349.

\bibitem{gellmann:67az}
M. Gell-Mann, {\it Recent work on representations of current
algebra}, International School of Physics E. Majorana (1967).

\bibitem{gellmann:67ay}
M. Gell-Mann, D. Horn and J. Weyers, Proceedings of the Heidelberg
Conference  on High Energy Physics and Elementary particles, (1967).

\bibitem{grodski:67yz}
I.T. Grodsky, M. Martinis and M. Swiecki, Physical Review Letters
{\bf 19} (1967) 332.

\bibitem{casalbuoni:ac1971}
R. Casalbuoni, R. Gatto and G. Longhi, Physical Review {\bf D3}
(1971) 1499.

\bibitem{grodsky:67ab}
I.T. Grodsky and R.F. Streater, Physical Review Letters {\bf 21}
(1967) 695.

\bibitem{casalbuoni:73ab}
R. Casalbuoni and G. Longhi, Il Nuovo Cimento {\bf 15A} (1973) 591.

\bibitem{streater:64ab}
R.F. Streater and A.S. Wightman, {\it PCT, Spin and Statistics, and
All That}, W.A. banjamin and Co. (1964).

\bibitem{abers:67ab}
E. Abers, I.T. Grodsky and R.E. Norton, Physical Review {\bf 159}
(1967) 1222.

\end{thebibliography}
\end{document}